\documentclass[aps,prb,twocolumn,raggedbottom,floatfix]{revtex4}
\usepackage{amsmath}
\usepackage{amssymb}
\usepackage{graphicx}

\newcommand{\eref}[1]{eqn.\ (\ref{#1})}
\newcommand{\Eref}[1]{Eqn.\ (\ref{#1})}
\newcommand{\fref}[1]{fig.\ \ref{#1}}
\newcommand{\sref}[1]{sec.\ \ref{#1}}

\newcommand{\cre}[1]{c^\dagger_{#1}}
\newcommand{\des}[1]{c^{\phantom\dagger}_{#1}}
\newcommand{\nhat}[1]{\hat n_{#1}}
\newcommand{\I}{\mathrm{i}}
\newcommand{\K}{\mathbf{k}}
\newcommand{\DeltaR}{\Delta_\mathrm{R}}
\newcommand{\DeltaI}{\Delta_\mathrm{I}}
\newcommand{\DeltaN}{\Delta_\mathrm{0}}
\newcommand{\dband}{d^\mathrm{b}}

\newcommand{\SigmaR}{\Sigma^\mathrm{R}}
\newcommand{\SigmaI}{\Sigma^\mathrm{I}}
\newcommand{\half}{\tfrac{1}{2}}
\newcommand{\PiNought}{\;{}^0\Pi}

\newcommand{\D}{\mathrm{d}}
\newcommand{\sgn}{\mathop{\mathrm{sgn}}}
\renewcommand{\Im}{\mathop{\mathrm{Im}}}

\newcommand{\PV}{\mathop{\mathrm{P}}}

\newcommand{\Z}{z}

\newcommand{\viz}{\emph{viz.\ }}
\newcommand{\ie}{\emph{i.e.\ }}
\newcommand{\eg}{\emph{e.g.\ }}

\begin{document}

\title{Anderson impurity model in a semiconductor}
\author{Martin R. Galpin}
\author{David E. Logan}
\affiliation{Physical \& Theoretical Chemistry Laboratory, Oxford University, South Parks Road, Oxford, OX1~3QZ, UK.}

\begin{abstract}
We consider an Anderson impurity model in which the locally correlated orbital is coupled 
to a host with a gapped density of states. Single-particle dynamics 
are studied, within a perturbative framework that includes both explicit second-order perturbation theory and self-consistent perturbation theory to all orders in the interaction. Away from particle-hole symmetry the system is shown to be a generalized Fermi liquid (GFL) in the sense of being perturbatively connectable to the non-interacting limit; and the exact Friedel sum rule for the GFL phase is obtained. We show by contrast that the particle-hole symmetric point of the model is not perturbatively connected to the non-interacting limit, and as such is a non-Fermi liquid for all non-zero gaps. Our conclusions are in agreement with NRG studies of the problem.
\end{abstract}

\pacs{72.15.Qm Scattering mechanisms and Kondo effect. 75.20.Hr Local moment in compounds and alloys; Kondo effect, valence fluctuations, heavy fermions}

\maketitle

\section{Introduction}

The Anderson impurity model (AIM), in which a single, locally correlated
orbital 
couples to a non-interacting metallic band of electrons, is a longstanding paradigm of strongly-correlated electron physics. Conceived originally\cite{anderson} to explain the formation of localized magnetic moments on impurities in non-magnetic hosts, it has since formed the cornerstone of our understanding of the Kondo effect\cite{hewsonbook} and related many-body phenomena. Interest in the area
is currently particularly strong, both experimentally and theoretically, after the Kondo effect was predicted\cite{nglee,raikh} and then directly
confirmed\cite{goldhaber,cronenwett} to arise in mesoscopic quantum dot systems.\cite{kouwenhovenqdreview} 

After some 50 years of intense theoretical work, the spin-$\half$ Kondo effect as manifest in Anderson's original model is naturally
rather well understood\cite{hewsonbook}. Below some characteristic Kondo temperature $T_\mathrm{K}$, a complex many-body state develops in which the impurity spin is completely screened by the host metal, leading at low energies to a `local' Fermi-liquid and
universal transport properties. 

Being a low-energy phenomenon, the Kondo effect is of course crucially dependent on both conduction band states near the Fermi level and the low-energy spin degrees of freedom of the impurity. This has inspired much research into other quantum impurity models---involving more complex impurities and/or host densities of states---with the aim of identifying the various types of Kondo effect that may arise, the conditions under which they do so, and the novel physics that results when Kondo screening cannot be achieved\cite{coxzawadowski,nrgreview}.

Here we consider the notionally simple problem of an Anderson impurity in a gapped host, where the density of states vanishes over a finite range about the chemical potential, a model not only of relevance to Anderson impurities in semiconductors but also\cite{sakaisuperconductor} to the topical issue of impurities in BCS superconductors\cite{impsupercondreview,bauersupercond}. In removing the all-important low-lying states of the host, one would certainly expect the Kondo effect to be precluded for large enough gaps: the question is, can the effect still arise for sufficiently-small gaps, or is it destroyed as soon as a gap is opened?

This question has indeed been the subject of a number of previous papers.
Poor man's scaling, the $1/N$ expansion and the non-crossing approximation predict \cite{saso, ogurasaso} that the Kondo effect always arises whenever the gap is less than the Kondo temperature in the absence of the gap, while for larger gaps the system undergoes a quantum phase transition to an `local moment' (LM) phase where the impurity spin remains unscreened as $T\to 0$. In addition the problem has been studied numerically by the density-matrix renormalization group\cite{yu} and quantum Monte Carlo\cite{saso, ogurasaso, take}, but with no general consensus reached regarding the nature of the quantum phase transition. The numerical renormalization group (NRG)\cite{chenjaya,take} on the other hand has been used to argue that the Fermi-liquid regime associated with the Kondo effect exists only away from particle--hole-symmetry, and then only below a certain critical gap. In the particle-hole--symmetric limit it is found\cite{chenjaya,take} that the Kondo effect \emph{never} arises and the ground state is the doubly-degenerate LM phase for arbitrarily small gaps.

In this paper we study the problem analytically, within a perturbative framework which includes both explicit second-order perturbation theory and self-consistent perturbation theory to all orders \`a la Luttinger~\cite{luttanprop}.
In addition to confirming the basic predictions of the NRG study~\cite{chenjaya}, our analysis provides a number of exact results, including the analogue of the Friedel sum rule, which serve 
as touchstones for approximate theories of the gapped AIM (GAIM). In a subsequent 
paper\cite{nextpaper}, we present a local moment approach\cite{mattdel_asym} to the problem, the results of which agree very well with the conclusions of the present work.

\section{Preliminaries}
\label{sec:prelim}

In standard notation the generic Anderson Hamiltonian\cite{hewsonbook} is
\begin{equation}
\hat H = \sum_{\K,\sigma}\epsilon_\K\nhat{\K\sigma} + \sum_\sigma\epsilon_\I\nhat{\I\sigma} + U\nhat{\I\uparrow}\nhat{\I\downarrow} + \sum_{\K,\sigma}(V_{\I \K}^{\phantom\dagger}\cre{\K\sigma}\des{\I\sigma} + \mathrm{h.c.})
\label{eq:h}
\end{equation}
where $\nhat{j\sigma}=\cre{j\sigma}\des{j\sigma}$ is the number operator for $\sigma$-spin electrons on `site' $j$ 
(with $j=\I$ referring to the impurity site and $j\equiv\K$ to the host band states). 
The first term in \eref{eq:h} thus describes the non-interacting host band, 
the second and third terms describe the impurity with onsite Coulomb interaction $U$, and the fourth term hybridises the two. For a symmetric host band, the particle-hole symmetric limit 
corresponds to the special point $\epsilon_\mathrm{i} =-U/2$ (where $\hat H$ is invariant under a particle-hole transformation).

The dynamics of the model will be obtained from the retarded Green function $G_{ij;\sigma}(\omega)$
\begin{equation}
G_{ij;\sigma}(t)=-\I\theta(t)\langle \{\des{i\sigma}(t),\cre{j\sigma}\}\rangle ,
\label{eq:gt}
\end{equation}  
differentiation of which
leads straightforwardly to its equations-of-motion \cite{hewsonbook}; from which the impurity-diagonal Green function in the non-interacting $(U=0)$ limit 
follows.
Its Fourier transform, denoted by $g_{\I\I;\sigma}(\omega)$, is
\begin{equation}
g_{\I\I;\sigma}(\omega) = \frac{1}{\Z-\epsilon_\I-\Delta(\omega)},
\label{eq:giiw}
\end{equation}
with $\Delta(\omega)$ the host-impurity hybridisation function
\begin{equation}
\Delta(\omega)=\sum_{\K} \frac{|V_{\I\K}|^2}{\Z-\epsilon_\K}
\label{eq:delta}
\end{equation}
and $\Z=\omega+\I\eta$ with $\eta=0^+$ a positive infinitesimal. The `full' and non-interacting Green functions are related in the usual way by Dyson's equation
\begin{equation}
G_{\I\I;\sigma}(\omega)^{-1} = g_{\I\I;\sigma}(\omega)^{-1} - \Sigma_{\I\I;\sigma}(\omega)
\label{eq:dyson}
\end{equation}
with $\Sigma_{\I\I;\sigma}(\omega)$ the conventional (single) self-energy.

It is
convenient below to exploit the analytic structure of the impurity Green 
functions and their constituent parts. Let
$F(z)$ be analytic on the real axis and in the upper half plane, tending to zero as $|z|\to \infty$. Then, with
$F(\omega) = F^\mathrm{R}(\omega) - \I F^\mathrm{I}(\omega)$
for real $\omega$, 
one has the well known
dispersion relation
\begin{equation}
F^\mathrm{R}(\omega) = \frac{1}{\pi}\PV\int_{-\infty}^\infty \D\omega' \frac{F^\mathrm{I}(\omega')}{\omega-\omega'}
\label{eq:generalfhilbert}
\end{equation}
(with  $\PV$ denoting a
principal value), and the spectral representation 
\begin{equation}
F(\omega) = \frac{1}{\pi}\int_{-\infty}^\infty \D\omega' \frac{F^\mathrm{I}(\omega')}{\omega-\omega'+\I\eta}.
\label{eq:generalfspectral}
\end{equation}
In particular, the full impurity Green function can be determined entirely from its spectral function,
\begin{equation}
D_{\I\I;\sigma}(\omega)=-\frac{1}{\pi}\Im G_{\I\I\;\sigma}(\omega).
\label{eq:ditog}
\end{equation}

\begin{figure}
\includegraphics{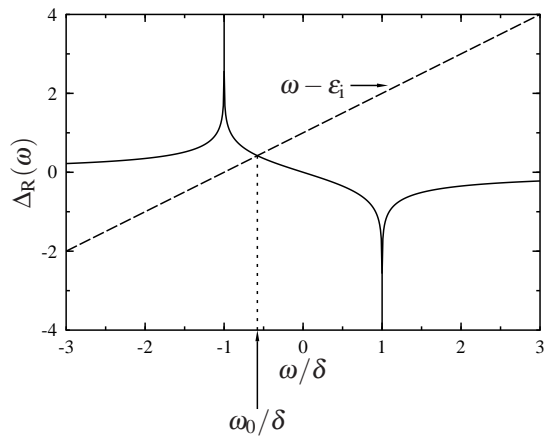}
\caption{\label{fig:delta}Real part of the hybridization function, $\Delta_\mathrm{R}(\omega)$, \emph{vs.} $\omega/\delta$ (solid line). The dashed line illustrates the solution of \eref{eq:dpolepos} as described in the text.}
\end{figure}

The results above are valid for whatever form the host takes in \eref{eq:h}: the details of the host bandstructure affect only the hybridisation function $\Delta(\omega)$. Assuming for simplicity that $V_{\I\K}=V$ is fixed, and writing $\Delta(\omega) = \DeltaR(\omega) - \I\DeltaI(\omega)$,
\eref{eq:delta} gives
\begin{equation}
\DeltaI(\omega) = \pi V^2\sum_\K\delta(\omega-\epsilon_\K) \equiv \pi V^2\rho(\omega)
\label{eq:deltairho}
\end{equation}
with $\rho(\omega)$ the host density of states. Hence, taking $\rho(\omega)$ to be constant except for a gap of full width $2\delta$ centred around $\omega=0$, we write $\DeltaI(\omega)$ for the \emph{gapped} Anderson model as 
\begin{equation}
\DeltaI(\omega) = 
\begin{cases}
0&\text{ for }|\omega|\le \delta\\
\DeltaN&\text{ for }|\omega| > \delta.
\end{cases}
\label{eq:deltagap}
\end{equation}
The corresponding $\DeltaR(\omega)$ ($=-\DeltaR(-\omega)$)
follows from the Hilbert transform
\eref{eq:generalfhilbert} as
\begin{equation}
\DeltaR(\omega) = -\frac{\DeltaN}{\pi}\ln\left|\frac{\omega+\delta}{\omega-\delta}\right|.\label{eq:deltargap}
\end{equation}
The logarithmic divergences of $\DeltaR(\omega)$ at $\omega=\pm\delta$ arising from the gap are shown in \fref{fig:delta}. 

The explicit form of the non-interacting spectrum $d_{\I\I;\sigma}(\omega)$ is readily obtained from eqns.\ (\ref{eq:giiw}), (\ref{eq:ditog}), (\ref{eq:deltagap}) and (\ref{eq:deltargap}).  Two distinct contributions arise: a continuum (or `band') part $\dband_{\I\I;\sigma}(\omega)$ arising when $\DeltaI(\omega)$ is non-zero 
\begin{equation}
\pi\DeltaN\dband_{\I\I;\sigma}(\omega) = \frac{\theta(|\omega|-\delta)\DeltaN^2}{\left[\omega-\epsilon_\I+\frac{\DeltaN}{\pi}\ln\left|\frac{\omega+\delta}{\omega-\delta}\right|\right]^2+\DeltaN^2},
\label{eq:dband}
\end{equation}
and a single pole inside the gap. 
This pole arises for all $\epsilon_\I$ when $\delta>0$, occuring at a frequency $\omega_0$ determined by solution of 
\begin{equation}
\omega_0-\epsilon_\I=\DeltaR(\omega_0).
\label{eq:dpolepos}
\end{equation}
That this equation always has one, and only one, solution is guaranteed by the monotonic decrease of $\DeltaR(\omega)$ from $\infty$ to $-\infty$ for $-\delta < \omega < \delta$, as seen from the construction in \fref{fig:delta}. Moreover, since $\Delta_\mathrm{R}(0)=0$, it is clear that $\omega_\mathrm{0}\lessgtr 0$ for $\epsilon_\mathrm{i}\lessgtr 0$, respectively. In the particle-hole symmetric limit, where $\epsilon_\mathrm{i}=0$, the pole lies precisely at $\omega_0=0$. We shall see that this zero-frequency pole is the 
basic reason for qualitatively distinct physics at particle-hole symmetry, compared to that elsewhere.

The weight of the pole, $Q_0$, follows straightforwardly from a Taylor expansion 
of $\DeltaR(\omega)$ as
\begin{equation}
Q_0 = \left[1+\frac{2\DeltaN}{\pi}\frac{\delta}{\delta^2-\omega_0^2}\right]^{-1},
\label{eq:dpoleweight}
\end{equation}
and hence the total non-interacting spectrum is given by
\begin{equation}
d_{\I\I;\sigma}(\omega) = \dband_{\I\I;\sigma}(\omega) + Q_0\delta(\omega-\omega_0).
\label{eq:dbp}
\end{equation}
This non-interacting Green function of course forms the basis for perturbation theory in $U$. We begin by analysing the problem to second-order in $U$, which is sufficient to highlight the essential features that lead to non-analyticities in the particle-hole symmetric limit.

\section{Second-order perturbation theory}
\label{sec:sopt}
\begin{figure}
\begin{center}
\includegraphics[scale = 0.8]{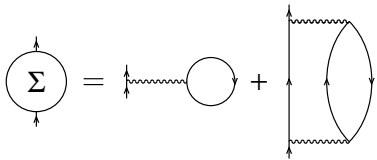}
\end{center}
\caption{\label{fig:2pt}Self-energy diagrams up to second order in the interaction $U$. The solid lines represent the non-interacting propagator, and the wavy lines represent $U$.}
\end{figure}

To second-order in $U$, the self-energy $\Sigma(\omega)$ ($=\Sigma^\mathrm{R}(\omega)-\mathrm{i}\Sigma^\mathrm{I}(\omega)$) is expanded diagramatically as shown in \fref{fig:2pt}. To simplify the notation, we henceforth drop the $\I\I$ subscripts, since we focus on the local impurity Green function. 
We also drop the spin index, as the Green functions obtained are independent of $\sigma$. 

Translating the diagrams in \fref{fig:2pt} using the Feynman rules for the time-ordered Green function gives 
\begin{equation}
\Sigma(\omega)=\half U n_0 + \Sigma_2(\omega)
\label{eq:sigmapt}
\end{equation}
where $n_0= 2\int_{-\infty}^0\D\omega\; d(\omega)$ is the impurity charge in the non-interacting limit,
\begin{equation}
\Sigma_2(\omega)=\frac{U^2}{2\pi\I}\int_{-\infty}^\infty\D\omega' \PiNought(\omega') g(\omega+\omega'),
\label{eq:sigtwo}
\end{equation}
and $\PiNought(\omega)$ is the polarisation `bubble' 
\begin{equation}
\PiNought(\omega)=\frac{\I}{2\pi}\int_{-\infty}^\infty\D\omega' g(\omega')g(\omega'-\omega).
\label{eq:pinought}
\end{equation}
Using the spectral representation of the Green function (see \eref{eq:generalfspectral}), $\Im\PiNought(\omega)$ is expressible entirely in terms of $d(\omega)$, \viz 
\begin{equation}
\frac{1}{\pi}\Im\PiNought(\omega)=\int_0^{|\omega|}\D\omega' d(\omega')d(\omega'-|\omega|).
\label{eq:impinought}
\end{equation}
Likewise, $\SigmaI_2(\omega)$ can be written as
\begin{equation}
\SigmaI_2(\omega) = U^2\int_0^{|\omega|}\D\omega' \Im\PiNought(\omega')~d(\omega-\sgn(\omega)\omega').
\label{eq:sigmatwod}
\end{equation}
From eqns.\ (\ref{eq:impinought}) and (\ref{eq:sigmatwod}), which hold also for the retarded Green function under consideration here, the second-order self-energy 
$\Sigma_2(\omega)$ can be constructed from the non-interacting spectrum $d(\omega)$ alone.  These integrals can of course be evaluated numerically for any given $\epsilon_\I$ and $\delta$ to obtain the full frequency dependence of $\Sigma_2(\omega)$.
This is not our aim here; instead, we focus on the low-frequency behaviour (which can be determined analytically), since it is this that contains the key physics of the problem.

\subsection{Away from particle-hole symmetry}
 The generic behaviour away from the particle-hole symmetric limit 
can be examined by focussing on some given $\epsilon_\I<0$; results for $\epsilon_\I>0$ follow from a straightforward particle-hole transformation. 

Given that the pole in $d(\omega)$ lies at a frequency $\omega_0 < 0$ for $\epsilon_\I< 0$ (see 
\sref{sec:prelim}), it is readily shown from eqns.\ (\ref{eq:dbp}) and (\ref{eq:impinought}) that
\begin{multline}
\frac{1}{\pi}\Im\PiNought(\omega) = \theta(|\omega|+\omega_0) ~Q_0 ~ \dband(|\omega|+\omega_0) 
\\+ \int_0^{|\omega|}\D\omega' \dband(\omega')\dband(\omega'-|\omega|).
\label{eq:impinoughta}
\end{multline} 
Since $\dband(\omega)$ has no spectral weight for $|\omega|<\delta$, the second term of \eref{eq:impinoughta} is zero for $|\omega|<2\delta$. The additional contribution from the first term of \eref{eq:impinoughta} also contains a gap for $|\omega|<\delta+|\omega_0|$, and since $|\omega_0|<\delta$ we can write 
\begin{equation} 
\Im\PiNought(\omega)=0\;\;\; : \;\;\;|\omega|<\delta+|\omega_0|.
\end{equation}
Using this in \eref{eq:sigmatwod}, similar arguments can be made for the low-frequency form of $\Sigma_2(\omega)$; it too has a gap around $\omega=0$, given by
\begin{equation} 
\label{eq:sigma2gap}
\SigmaI_2(\omega)=0\;\;\; : \;\;\;-\delta-2|\omega_0|<\omega<2\delta+|\omega_0|.
\end{equation}

The gap in $\SigmaI_2(\omega)$ enables one to deduce the salient behaviour
of $\SigmaR_2(\omega)$. By writing $\SigmaI_2(\omega) \equiv \theta(|\omega|-\delta)\SigmaI_2(\omega)$, the Hilbert transform \eref{eq:generalfhilbert} gives
\begin{equation}
\SigmaR_2(\omega)=\frac{1}{\pi}\int_{-\infty}^{\infty}\D\omega'\frac{\theta(|\omega'|-\delta)\SigmaI_2(\omega')}{\omega-\omega'}
\end{equation}
and hence 
\begin{multline}
\SigmaR_2(\omega_1)-\SigmaR_2(\omega_2)=\\\frac{\omega_2-\omega_1}{\pi}\int_{-\infty}^{\infty}\D\omega'\left[\frac{\theta(|\omega'|-\delta)}{(\omega_1-\omega')(\omega_2-\omega')}\right]\SigmaI_2(\omega')
\end{multline}
If both $\omega_1$ and $\omega_2$ are within the gap, the term in square brackets above is non-negative for all $\omega'$. Since $\SigmaI_2(\omega)$ is necessarily non-negative, it follows that
\begin{equation}
\SigmaR_2(\omega_1)>\SigmaR_2(\omega_2)\;\;\; : \;\;\;-\delta < \omega_1<\omega_2 < \delta.
\label{eq:sigrmono}
\end{equation}
This monotonicity is important in
arguments below. 

Finally, the low-energy behaviour of $\Sigma_2(\omega)$ can be inserted into the Dyson equation, \eref{eq:dyson}, to obtain the impurity Green function to second-order in $U$. For $|\omega|<\delta$, 
\begin{equation}
G_2(\omega)\overset{|\omega|<\delta}=\frac{1}{\Z-\epsilon_\I-\half U n_0 -\DeltaR(\omega) - \SigmaR_2(\omega)}
\end{equation}
and, by using similar arguments to those in the non-interacting limit, the monotonicity of $\SigmaR_2(\omega)$ in \eref{eq:sigrmono} guarantees that $G_2(\omega)$ has a single pole at a frequency $\omega_{\mathrm{p}2}$ given by
\begin{equation}
\omega_{\mathrm{p}2}-\epsilon_\I-\half U n_0=\DeltaR(\omega_{\mathrm{p}2})+\SigmaR_2(\omega_{\mathrm{p}2}).
\end{equation}

Inside the gap the second-order Green function $G_2(\omega)$ is thus seen to have the same basic structure as the non-interacting $g(\omega)$. In particular, the $\omega\to 0$ quasiparticle behaviour of the Green function is obtained by
expanding $\SigmaR_2(\omega)$ about $\omega=0$,
\begin{equation}
G_2(\omega)\overset{\omega\to 0}\sim\frac{1}{{\Z}/{Z_2}-\epsilon_\I^*-\DeltaR(\omega)}
\end{equation}
where $\epsilon_\I^*=\epsilon_\I+\half U n_0 + \SigmaR_2(0)$ is the renormalized level energy, and
\begin{equation}
Z_2 = \left[1-\left(\frac{\partial \SigmaR_2(\omega)}{\partial \omega}\right)_{\omega=0}\right]^{-1}
\end{equation}
is the quasiparticle weight. That $G_2(\omega)$ is a renormalized version of the non-interacting
$g(\omega)$ at low-frequencies is of course a direct reflection of adiabatic continuity to the non-interacting limit, in which sense the system is a generalized Fermi liquid (GFL). For some fixed $\epsilon_\I$ and $\delta$, on switching on a small $U$, one expects the system to evolve smoothly into this perturbative state and as such to behave as a (local) Fermi liquid. As we now show, however, such behaviour does \emph{not} arise at the particle-hole symmetric limit of the model.

\subsection{The particle-hole symmetric limit}
That the behaviour at the particle-hole symmetric point is qualitatively different
from that described above, arises because (see \sref{sec:prelim}) the pole in the non-interacting
$d(\omega)$ lies precisely at $\omega_0=0$. On substituting \eref{eq:dbp} into \eref{eq:impinought}, $\Im\PiNought(\omega)$ itself contains a pole at zero frequency:
\begin{multline}
\label{eq:impinoughtph}
\frac{1}{\pi}\Im\PiNought(\omega)= \half Q_0^2\delta(\omega) \\+ Q_0\dband(\omega) + \int_0^{|\omega|}\D\omega' \dband(\omega')\dband(\omega'-|\omega|).
\end{multline} 
Physically, this pole describes zero-energy spin-flip excitations of the impurity at the non-interacting level: the $\omega =0$ pole in $d(\omega)$ 
reflects a state which is equally likely to contain a single $\uparrow$- or $\downarrow$-spin electron. 

The second-order self energy term $\SigmaI_2(\omega)$ is obtained as before by substituting \eref{eq:impinoughtph} into \eref{eq:sigmatwod}. Whereas away from particle-hole symmetry the key result was a \emph{gap} in $\SigmaI_2(\omega)$ around the Fermi level (see \eref{eq:sigma2gap}), here instead the pole in $\Im\PiNought(\omega)$ leads to a corresponding \emph{pole} in the self-energy:
\begin{equation}
\label{eq:sigma2iph}
\SigmaI_2(\omega)\overset{|\omega|<\delta}\sim \tfrac{1}{4} U^2 Q_0^3~\delta(\omega).
\end{equation}
Such behaviour is strikingly different from the GFL physics described above. The real part of $\SigmaR_2(\omega)$ obtained from \eref{eq:sigma2iph} diverges as $\omega\to 0$, i.e.
\begin{equation}
\SigmaR_2(\omega)\overset{\omega\to 0}\sim \frac{U^2 Q_0^3}{4\omega}.
\end{equation}
And the corresponding $G_2(\omega)$ 
$=[g(\omega)^{-1}-\tfrac{1}{2}Un_{0}-\Sigma_2(\omega)]^{-1}$
is thus of form
\begin{equation}
\label{eq:g2ph}
G_2(\omega)\overset{|\omega|<\delta}\sim\frac{1}{\Z - \DeltaR(\omega) - \frac{U^2 Q_0^3}{4\Z}}
\end{equation}
(noting that the renormalized level energy $\epsilon_\I^* $, given generally by
$\epsilon_\I^*=\epsilon_\I+ \SigmaR(0)$, vanishes by symmetry at the particle-hole symmetric
point $\epsilon_\I = -U/2$).

\Eref{eq:g2ph} clearly cannot be written as a renormalized version of the non-interacting 
$g(\omega)$, indicating that the particle-hole symmetric point is not perturbatively connected in $U$ to the non-interacting limit. 
Instead, it is readily shown by expanding the hybridisation function $\DeltaR(\omega)$ about $\omega=0$ that
\begin{equation}
\label{eq:g2ral}
G_2(\omega)\overset{\omega\to 0}\sim\frac{Q_0}{\Z - {\tfrac{1}{4} U_\mathrm{eff}^2}/{\Z}}
\end{equation}
with a renormalized interaction $U_\mathrm{eff} = U Q_0^2$ (and $Q_{0} = [1+\frac{2\Delta_{0}}{\pi\delta}]^{-1}$ from \eref{eq:dpoleweight}).
The significance of this result is that it
is a renormalized version of the \emph{atomic limit} ($V=0$) propagator, 
$G_\mathrm{AL}(\omega) = [\Z - \tfrac{1}{4}U^2/\Z]^{-1}$, indicative of the LM nature of the phase arising at particle-hole symmetry. 
Instead of the single pole inside the gap seen away from particle-hole symmetry, \emph{two} poles thus arise, at $\omega = \pm \frac{1}{2}U_\mathrm{eff}$, and representing as such
renormalized versions of the atomic limit Hubbard satellites (themselves occurring at
$\omega = \pm \frac{1}{2}U$).

Even at the basic level of second-order perturbation theory, one sees then that introducing a gap in the conduction band of the Anderson model changes significantly its low-energy physics, with non-analyticities arising in the perturbative self-energies in the particle-hole symmetric limit. We now extend the treatment to arbitrary order in the interaction $U$, bolstering the conclusions drawn above and in doing so obtaining 
some exact results for the problem.

\section{Infinite order perturbation theory in $U$}
\label{sec:inforder}
One cannot of course derive perturbatively the full frequency dependence of the impurity Green function to all orders, but its behaviour inside the gap can be ascertained exactly using some simple arguments. Our main result is to determine the conditions under which perturbation theory in $U$ is valid, and the resulting low-frequency behaviour of the impurity Green function.

\begin{figure}
\includegraphics[scale = 0.8]{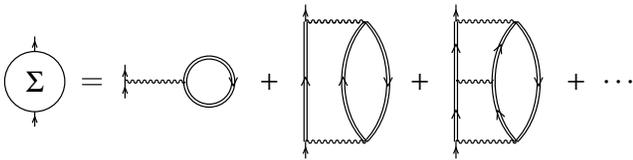}
\caption{\label{fig:selfskel}Skeleton expansion of the self-energy. The exact $\Sigma(\omega)$ is given by the sum of all skeleton self-energy diagrams constructed from the exact $G(\omega)$.}
\end{figure}

The argument follows that of the classic paper by Luttinger\cite{luttanprop}, in which the skeleton expansion of the self-energy
is used to obtain self-consistently the low-frequency behaviour of a wide class of 
interacting problems to all orders in the interaction. The basis of the approach is that if perturbation theory holds, the exact self-energy is equal to the sum of all skeleton self-energy diagrams constructed from the exact Green function, as shown diagramatically in \fref{fig:selfskel}. By analysing the general low-frequency behaviour of all such skeleton diagrams, we show that this condition is satisfied 
only if the exact Green function contains a \emph{single} pole inside the gap at a non-zero frequency.

A key insight of ref. \onlinecite{luttanprop} is that the low-energy behaviour of the self-energy can be deduced perturbatively---\emph{via} the skeleton expansion---from the low-energy single-particle excitations of the exact $G(\omega)$. To be more precise, we refer to the positive frequency excitations of $D(\omega)$ as `particle energies' and the negative frequency excitations as `hole energies'. Then, it can be shown\cite{luttanprop} quite generally that $\Im\Sigma(\omega)$ has weight at positive frequencies corresponding to $n+1$ particle energies minus $n$ hole energies, or at negative frequencies corresponding to $n+1$ hole energies minus $n$ particle energies, with $n\ge 1$. Combining this result with the Dyson equation [giving 
$G(\omega)$ in terms of $\Sigma(\omega)$] generates a pair of self-consistency equations for $\Sigma(\omega)$. In what follows, we show that any self-energy that contains poles inside the gap (\ie for $|\omega|<\delta$) is inconsistent with these 
requirements.

\begin{figure}
\includegraphics{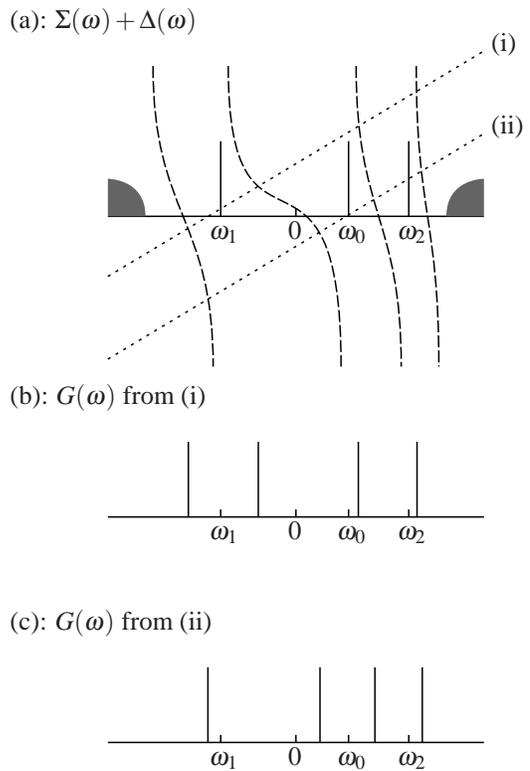}
\caption{\label{fig:poles}Sketch showing how poles in $\Sigma(\omega)$ lead (via the Dyson equation) to poles in $G(\omega)$. (a) Solid lines represent the schematic form of $\SigmaI(\omega)+\DeltaI(\omega)$ with poles in the gap, dashed lines represent the corresponding $\SigmaR(\omega)+\DeltaR(\omega)$ from Hilbert transformation, and the dotted lines show the constructions used for determing the poles in $G(\omega)$ from Dyson's equation (\ref{eq:dyson}). (b) The pole structure of $G(\omega)$ obtained from case (i) of (a). (c) The pole structure of $G(\omega)$ obtained from case (ii) of (a).}
\end{figure}

Consider the situation in which $\SigmaI(\omega)$ contains at least one pole in the range $-\delta < \omega < \delta$, and label by $\omega_0$ the frequency of the pole closest to $\omega=0$. In the following 
we take $\omega_0 \ge 0$; similar reasoning can be applied if the closest pole is at a negative frequency. Then define $\omega_1$ to be the least negative frequency at which $\SigmaI(\omega)$ has weight (either in the form of another pole, or the upper edge of the negative-frequency continuum), and $\omega_2$ to be the next frequency above $\omega_0$ at which $\SigmaI(\omega)$ has weight. We sketch the resulting $\SigmaI(\omega)+\DeltaI(\omega)$ in \fref{fig:poles}(a), taking (without loss of generality) both $\omega_1$ and $\omega_2$ to correspond to discrete poles rather than the edges of the continua.

In \fref{fig:poles}(a), we also sketch the $\SigmaR(\omega)+\DeltaR(\omega)$ corresponding to the $\SigmaI(\omega)+\DeltaI(\omega)$ described above. Its form follows from the Hilbert transform \eref{eq:generalfhilbert}. Within each frequency range in which $\SigmaI(\omega)$ is zero, an argument akin to that before \eref{eq:sigrmono} shows that $\SigmaR(\omega)$ is monotonically decreasing. Moreover, at frequencies where $\SigmaI(\omega)$ has poles, it is readily seen from \eref{eq:generalfhilbert} that $\SigmaR(\omega)$ diverges as $1/(\omega -\omega_\mathrm{p})$ (with $\omega_\mathrm{p}$ the particular pole frequency). Hence $\SigmaR(\omega)+\Delta(\omega)$ spans the full range from $\infty$ to $-\infty$ between each excitation in $\SigmaI(\omega) + \DeltaI(\omega)$.

The real and imaginary parts of the schematic self-energy can then be inserted into the Dyson equation to determine $G(\omega)$. Since $\DeltaI(\omega) > 0$ for $|\omega| > \delta$, it follows from eqns. (\ref{eq:giiw}) and (\ref{eq:dyson}) that $G(\omega)$ has poles only for $|\omega|<\delta$, at frequencies given by the solution of $\omega-\epsilon_\I = \SigmaR(\omega) + \DeltaR(\omega)$. 

The resultant poles in $G(\omega)$ thus lie at the frequencies where the dashed lines in \fref{fig:poles} intersect straight lines of unit slope, as shown by the dotted lines in \fref{fig:poles}(a). From the diagram, it is clear that within the region $\omega_1<\omega<\omega_2$ there will always be two such poles in $G(\omega)$. Either there is one on each side of $\omega=0$, which we denote by `case (i)' and show schematically in \fref{fig:poles}(b); or both poles are at $\omega>0$, denoted `case (ii)' (see \fref{fig:poles}(c)) [Note that if $\omega_0=0$, then the only possible case is (i)].

Now consider the $\omega>0$ behaviour of $\SigmaI(\omega)$ obtained by inserting this $G(\omega)$ into the skeleton expansion. As discussed above, this is obtained~\cite{luttanprop}
by considering the energies of possible particle and hole excitations in $G(\omega)$. In case 
(i), the lowest particle energy is at an $\omega>\omega_0$, and the lowest hole energy is at an $\omega\le 0$. Since\cite{luttanprop} the lowest-energy pole in $\SigmaI(\omega)$ for $\omega>0$ is at an energy corresponding to two particle energies minus one hole energy, this lowest-energy pole is necessarily at a frequency $\omega > 2\omega_0$ \emph{and is therefore not consistent with the starting $\SigmaI(\omega)$ in \fref{fig:poles}(a)}. Likewise in case (ii), the lowest particle energy is at an $\omega\ge 0$, and the lowest hole energy is at an $\omega<\omega_1$. It follows that any resulting pole in $\SigmaI(\omega)$ is at an $\omega > |\omega_1|$ with $|\omega_1|\ge\omega_0$ by definition. Hence case (ii) is also not self-consistent.

We have thus shown that there are no self-consistent perturbative solutions of the GAIM with a pole in $\SigmaI(\omega)$ at an $\omega_0 \ge 0$. Similar arguments can be used to prove the same is true for a pole at $\omega_0<0$. Therefore, \emph{any perturbative (Fermi liquid) solution cannot contain poles in $\SigmaI(\omega)$ inside the gap}. 

Finally we show that given the conclusion drawn above, particle-hole symmetry \emph{always} leads to a breakdown of perturbation theory.
To that end, imagine starting with a perturbative $\SigmaI(\omega)$ [i.e. no poles in the gap] in the particle-hole symmetric limit. Its corresponding real part is monotonically decreasing inside the gap, and furthermore satisfies $\epsilon_\I + \SigmaR(0) = 0$ by particle-hole symmetry. Therefore the Green function obtained by substituting this self-energy into the Dyson equation (\ref{eq:dyson}) must have a pole at $\omega=0$. But then the $\SigmaI(\omega)$ obtained by substituting this $G(\omega)$ into the skeleton expansion would itself have a pole at zero frequency, breaking the perturbative self-consistency of the skeleton expansion. \emph{Hence the particle-hole symmetric point cannot be perturbatively connected to the non-interacting limit, and as such is a non Fermi-liquid, for all $\delta > 0$.}

Away from particle-hole symmetry by contrast, the pole in $G(\omega)$ arising from a perturbative $\SigmaI(\omega)$ can (and will) be at a non-zero frequency. In this case the skeleton expansion would not lead to a pole inside the gap in $\SigmaI(\omega)$, which is consistent with the form of the starting $\SigmaI(\omega)$. Hence such perturbative solutions can in principle exist; and
the results of \sref{sec:sopt} naturally lead us to expect them, at least for some non-vanishing
radius of convergence in $U$.

\section{The Luttinger-Ward functional}
By closing the skeleton self-energy diagrams in \fref{fig:selfskel} with an additional propagator line, one obtains the Luttinger-Ward functional $\Phi[\mathcal{G}]$ \cite{agdbook}, from which the exact self-energy follows from the functional derivative
\begin{equation}
\Sigma = \left(\frac{\delta \Phi[\mathcal{G}]}{\delta \mathcal{G}}\right)_{\mathcal{G}=G}.
\label{eq:sigphi}
\end{equation}

Using \eref{eq:sigphi}, and that $\Phi[\mathcal{G}]$ is invariant\cite{agdbook} under a frequency shift of all its propagators $\mathcal{G}(\omega)$, it can be shown that the  `Luttinger integral' $I_\mathrm{L}$ must vanish:
\begin{equation}
I_\mathrm{L} = \Im\int_{-\infty}^0 \D\omega \frac{\partial \Sigma(\omega)}{\partial \omega} G(\omega) = 0.
\label{eq:il}
\end{equation}
This can be used in turn to derive a generalized Friedel sum rule\cite{hewsonbook} for the GAIM, which must be satisfied by any perturbative solution of the problem.
Consider the change in the density of states upon introducing the impurity, $\Delta\rho(\omega) = \rho(\omega) - \rho_\mathrm{host}(\omega)$. This is given by
\begin{equation}
\label{eq:lw1}
\Delta\rho(\omega)=-\frac{1}{\pi}\Im\Biggl[\sum_\K G_{\K \K}(\omega) + G(\omega) - \sum_{\K}\frac{1}{\Z-\epsilon_\K}\Biggr].
\end{equation}
The equations of motion for the host Green functions generate \cite{hewsonbook}
\begin{align}
\sum_\K G_{\K \K}(\omega) &= \sum_{\K}\frac{1}{\Z-\epsilon_\K} + \sum_{\K}\frac{|V_{\I\K}|^2}{(\Z-\epsilon_\K)^2} G(\omega)\\
&= \sum_{\K}\frac{1}{\Z-\epsilon_\K} - \frac{\partial \Delta(\omega)}{\partial \omega} G(\omega),
\end{align}
and the Dyson equation (\ref{eq:dyson}) gives
\begin{equation}
\label{eq:lw3}
\frac{\partial}{\partial \omega} G(\omega)^{-1} = 1 - \frac{\partial \Delta(\omega)}{\partial \omega} -\frac{\partial \Sigma(\omega)}{\partial \omega}.
\end{equation}
Combining eqns. (\ref{eq:lw1})--(\ref{eq:lw3}) thus leads to the general result \cite{hewsonbook}
\begin{equation}
\label{eq:lw4}
\Delta\rho(\omega)= - \frac{1}{\pi}\Im\left[\frac{\partial\ln G(\omega)^{-1}}{\partial \omega} + \frac{\partial \Sigma(\omega)}{\partial \omega} G(\omega)\right].
\end{equation} 
The number of electrons introduced by the impurity, $n_\mathrm{imp}$, is now obtained by integrating $2\Delta\rho(\omega)$ up to $\omega=0$ (the factor of $2$ arising from a spin sum). From \eref{eq:lw3}, and using the Luttinger integral condition \eref{eq:il}, one has
\begin{equation}
\label{eq:lw5}
n_\mathrm{imp} = -\frac{2}{\pi}\Im\int_{-\infty}^0 \D\omega \frac{\partial}{\partial \omega} \ln G(\omega)^{-1}.
\end{equation}
From equation (\ref{eq:dyson}) it follows that
\begin{equation}
\Im\ln G(\omega)^{-1} = \frac{\pi}{2}-\tan^{-1}\left[\frac{\omega-\epsilon_\I-\DeltaR(\omega)-\SigmaR(\omega)}{\DeltaI(\omega)+\SigmaI(\omega)+\eta}\right]
\end{equation}
(taking the principal range $-\pi/2 < \tan^{-1}x < \pi/2$), and hence \cite{hewsonbook}
\begin{equation}
\label{eq:wilber}
n_\mathrm{imp} = 1-\frac{2}{\pi}\tan^{-1}\left[\frac{\epsilon_\I+\DeltaR(0)+\SigmaR(0)}{\DeltaI(0)+\SigmaI(0)+\eta}\right].
\end{equation}

This Friedel sum rule is quite general for the Anderson impurity model. For the GAIM in particular one has $\DeltaI(0)=0$ $=\DeltaR(0)$, and the results for the self-energy deduced in the previous section showed that 
$\SigmaI(0)$ necessarily vanishes if perturbation theory holds (GFL phase). Hence
$n_\mathrm{imp} = 1-\frac{2}{\pi}\tan^{-1}\left[\frac{\epsilon_\I+\SigmaR(0)}{\eta}\right]$,
from which
\begin{equation}
\label{eq:nimplast}
n_\mathrm{imp}= 1-\sgn(\epsilon_\I^*)
\end{equation} 
with $\epsilon_\I^* = \epsilon_\I+\SigmaR(0)$ the renormalized level energy.
For \emph{any} $\epsilon_\I^* <0$, the additional number of electrons induced by the
impurity is thus $n_\mathrm{imp} =2$, while $n_\mathrm{imp} =0$ for any 
$\epsilon_\I^* >0$. This behaviour---exclusively integral values of $n_\mathrm{imp}$---is  physically natural for a system with a gapped spectrum,
although it is 
of course in marked contrast to the metallic
(gapless) Anderson model for which $\DeltaI(0)>0$ and hence (from \eref{eq:wilber})
$n_\mathrm{imp}$ is a continuous function of $\epsilon_\I^*$.

\section{Conclusion}
Finally, we tie together the results of the previous sections. We have shown that self-consistent perturbative solutions of the problem must necessarily have a non-zero renormalized level, $\epsilon_\I^*$, since this leads to a single pole in $G(\omega)$ at a non-zero frequency and hence a perturbatively-consistent $\Sigma(\omega)$ from the skeleton expansion. If $\epsilon_\I^*$ is negative the pole in $G(\omega)$ is at a negative frequency and 
$n_\mathrm{imp} = 2$; while if $\epsilon_\I^*$ is positive the pole is at a positive frequency and $n_\mathrm{imp} = 0$.

In \sref{sec:inforder}, we deduced that such perturbative solutions of the problem can only exist away from particle-hole symmetry: it is impossible to construct a $G(\omega)$ that is both consistent with particle-hole symmetry \emph{and} the perturbative skeleton expansion of the self-energy. Hence the particle-hole--symmetric point of the problem is a non-Fermi-liquid LM phase, and any approximation to the GAIM must therefore take this into account\cite{nextpaper}. Away from particle-hole symmetry, there is nothing to prevent the existence of perturbative GFL solutions, and these are indeed found numerically in \eg the NRG\cite{chenjaya}. 

Finally we note that since the derivation of the generalized Friedel sum rule, \eref{eq:nimplast} above, has been obtained from a perturbative construction of the Luttinger Ward functional $\Phi[\mathcal{G}]$, it is not valid in the LM phase. At the particle-hole--symmetric point, it is readily shown instead that $n_\mathrm{imp}=1$ by symmetry. In our forthcoming local moment approach to the model\cite{nextpaper}, we argue that $n_\mathrm{imp}=1$ is a general result that also holds for the LM phase away from particle-hole symmetry, and present results 
of this non-perturbative approach for the dynamics of the problem in the two phases. \\

\begin{acknowledgments}
We are grateful to the EPSRC for financial support.
\end{acknowledgments}


\end{document}